\documentclass{article}

\usepackage{arxiv}
\usepackage{amsmath}
\usepackage[utf8]{inputenc} % allow utf-8 input
\usepackage[T1]{fontenc}    % use 8-bit T1 fonts
\usepackage{hyperref}       % hyperlinks
\usepackage{url}            % simple URL typesetting
\usepackage{booktabs}       % professional-quality tables
\usepackage{amsfonts}       % blackboard math symbols
\usepackage{nicefrac}       % compact symbols for 1/2, etc.
\usepackage{microtype}      % microtypography
\usepackage{graphicx}
\usepackage{doi}
\usepackage{xcolor}

\title{The theoretical basis of\\ reservoir pressure in arteries}

\author{Kim H. Parker, \\
	Department of Bioengineering\\
	Imperial College \\
	London SW7 2AZ, UK \\
        ORCID ID: 0000-0001-6971-0682 \\
	\texttt{k.parker@imperial.ac.uk}
 \and
 Alun D. Hughes\\
	Institute of Cardiovascular Science\\
	University College London\\
        London WC1E 6BT \\
        ORCID ID: 0000-0001-5432-5271 \\
        \texttt{alun.hughes@ucl.ac.uk}
 }

\renewcommand{\shorttitle}{The theoretical basis of reservoir pressure}

\hypersetup{
pdftitle={A template for the arxiv style},
pdfsubject={q-bio.NC, q-bio.QM},
pdfauthor={David S.~Hippocampus, Elias D.~Striatum},
pdfkeywords={First keyword, Second keyword, More},
}

\begin{document}
\maketitle
\pagestyle{myheadings}
\markright{The theoretical basis of reservoir pressure}

\begin{abstract}
	The separation of measured arterial pressure into a reservoir pressure and an excess pressure was introduced nearly 20 years ago as an heuristic hypothesis. Since then it has gained some traction through a number of retrospective epidemiological studies that show that various measures of the reservoir and excess pressure are independent risk factors for the development of cardiovascular disease. 

    We demonstrate that a two-time asymptotic analysis of the 1-D conservation equations in each artery coupled with the separation of the smaller arteries into inviscid and resistance arteries, based on their resistance coefficients, results, for the first time, in a formal derivation of the reservoir pressure. 
	
	The key to the two-time analysis is the existence of a fast time associated with the propagation of waves through the arteries and a slow time associated with the convective velocity of the blood. The ratio between these two time scales is given by the Mach number; the ratio of a characteristic convective velocity to a characteristic wave speed. If the Mach number is small, a formal asymptotic analysis can be carried out which is accurate to the order of the square of the Mach number.

    The slow-time conservation equations involve a resistance coefficient that models the effect of viscosity on the convective velocity. On the basis of this resistance coefficient, we separate the arteries into the larger \textit{inviscid} arteries where the coefficient is negligible and the smaller \textit{resistance} arteries where it it is not negligible. The slow time pressure in the inviscid arteries is shown to be spatially uniform but varying in time. We define this pressure as the reservoir pressure. Dynamic analysis using mass conservation in the inviscid arteries shows that the reservoir pressure accounts for the storage of potential energy by the distension of the elastic inviscid arteries during early systole and its release during late systole and diastole. This analysis thus provides a formal derivation of the reservoir pressure and its physical meaning.
    
\end{abstract}

% keywords can be removed

\section{Introduction}

Reservoir pressure was introduced to arterial hemodynamics as part of an heuristic hypothesis, subsequently termed the reservoir-wave hypothesis.\cite{Wang2003} In that paper the reservoir pressure was called the Windkessel pressure and was calculated from locally measured pressure and velocity using the classical Windkessel theory.\cite{Frank1899}

In a follow up paper, it was recognised that the pressure in Windkessel theory is presumed to be uniform throughout the arterial compartment while the local 'Windkessel pressure' was observed to propagate along the aorta at the local wave speed. The hypothesised pressure was renamed the reservoir pressure to denote this difference and the reservoir-wave hypothesis was reformulated to account for the observed propagation of the reservoir pressure down the aorta.\cite{Tyberg2005} 

The reservoir-wave hypothesis in its most general form asserts that 'it may be useful to separate arterial pressure into two components: the reservoir pressure and the excess pressure. The reservoir pressure waveform is assumed to be the same in every artery but delayed by the wave travel time from the aortic root. The excess pressure is defined as the difference between the measured pressure and the reservoir pressure and varies both in time and space.' 

Mathematically this can be expressed
\begin{equation}
\label{eq:Pr}
P(x,t) = P_r(t-\delta t(x)) + P_e(x,t)
\end{equation}
where $P$ is the measured pressure, $P_r$ is the reservoir pressure, $P_e$ is the excess pressure and $x$ and $t$ are the spatial and temporal variables. $\delta t$ is a time delay that depends on the distance from the aortic root. An analysis using the calculus of variations showed that the reservoir pressure, so defined, generated the minimum hydraulic work for any given flow waveform.\cite{Parker2012}

The Windkessel model of arterial hemodynamics, despite extensive efforts by Frank and his followers, does not describe arterial pressure during systole with enough precision to be useful clinically and consequently is generally seen by the cardiovascular community as only of historical interest. During diastole, however, the Windkessel model provides an excellent description of arterial pressure, particularly in the elderly; the time constant of the observed exponential fall-off of pressure $RC$, where $R$ is the net arterial resistance and $C$ is the net compliance of the arteries, is  measured clinically, particularly in the pulmonary circulation, and is frequently cited as one of the descriptors of the performance of the cardiovascular system. 

The reservoir-wave hypothesis is an attempt to retain the ability of the Windkessel model to describe diastolic hemodynamics while positing the excess  pressure to explain the difference between measured arterial pressure and the reservoir pressure during systole. When the hypothesis was first formulated, we were unable to provide theoretical grounds for the existence of the reservoir pressure and were careful to describe it as an heuristic hypothesis.

The concept of a reservoir pressure elicited some controversy amongst cardiovascular theoreticians. \cite{Westerhof2015}\cite {Mynard2014} Nevertheless, a number of clinical cardiologists and epidemiologists have  carried out studies to test the utility of the concept (see for example,\cite{Hametner2014} \cite{Narayan2015},\cite{Behnam2019},\cite{Aizawa2021}). The studies are too diverse to summarise in detail but generally it has been found that parameters associated with the reservoir pressure, such as its peak value or area under the curves of reservoir or excess pressure, are significant indicators of the risk of cardiovascular disease that are independent of other risk factors.

Encouraged by these results, we explore the theoretical and physical basis of the reservoir and excess pressures.

\section{Theoretical background}

This analysis is based on the two  fundamental principles of continuum mechanics: the conservation of mass and momentum. The conservation of mass is one of the basic tenets of non-relativistic physics which, coupled with the near incompressibility of blood, can be expressed as a conservation of the volume of blood.
The conservation of momentum is the application of Newton's second law to the control volume that is central to the Eulerian analysis of fluid mechanics.

When applied to incompressible flow in a uniform compliant tube, the conservation principles and the constitutive equation relating area and pressure can be written as two partial differential equations for the pressure $P$ and mean velocity $U$.\cite{Parker2009} Using subscript notation to denote partial derivatives in terms of the independent variables, the axial distance $x$ and the time $t$
\begin{align}
    P_t + UP_x + \rho c^2U_x &= 0 \\
    U_t + \frac{1}{\rho}P_x + UU_x &= -\mathcal{R}U
\end{align}
where $\rho$ is the density of blood, $c^2 = \frac{A}{\rho} \frac{dP}{dA}$ and $\mathcal{R}$ is a resistance coefficient. For Poiseuille flow, $\mathcal{R} = \frac{8\pi \mu}{\rho A}$, where $\mu$ is the coefficient of viscosity and $A$ is the cross-sectional area of the tube. The parameter $c$, derived by the the method of characteristics, is the wave speed of the axial distension wave that dominates arterial hemodynamics. The 1-D equations (2 \& 3) are based on a number of approximations which are discussed in detail in previous work.\cite{Parker2009}  These equations are the basis of Wave Intensity Analysis which describes arterial hemodynamics in terms of forward and backward travelling wavelets.\cite{Parker2009}

\subsection{Two-time analysis}

In this section we introduce a new analysis based on a two-time asymptotic analysis to justify the separation of $P$ into two components. The analysis takes advantage of the large difference between the wave speed $c$ and a characteristic convective velocity $U$ which is usually expressed in terms of the Mach number, $m = \frac{U}{c} \ll 1$. 

It is most convenient in asymptotic analysis to deal with non-dimensional equations. Defining the characteristic distance $X^*$, characteristic velocity $U^*$, characteristic time $\frac{X^*}{U^*}$, characteristic wave speed $c^*$ and characteristic pressure $P^* = \rho c^* U^*$, we obtain the non-dimensional form of the conservation equations 
\begin{align}\label{eq:non-dim_cons_eqs}
    p_t + up_x + \frac{1}{m}u_x &= 0\\
    u_t + \frac{1}{m}p_x + uu_x &= -ru
\end{align}
where $p=\frac{P}{P^*}$, $u=\frac{U}{U^*}$ and $r = \frac{x^*\mathcal{R}}{U^*}$ is the non-dimensional resistance coefficient. For convenience we have retained the same symbols for dimensional and non-dimensional distance and time.

In the two-time analysis we assume that there are two distinct times; a slow time $T=t$ commensurate with convection and a fast time $\tau=\frac{1}{m}t$ commensurate with wave travel times. By the chain rule for partial derivatives, the partial derivative wrt $t$ can be written
\begin{equation*}
    \frac{\partial}{\partial t} =
    \frac{\partial}{\partial T} +
    \frac{1}{m}\frac{\partial}{\partial \tau}
\end{equation*}
Applying this to the time derivatives in the non-dimensional conservation equations and separating the orders of magnitude we obtain to $O(m)$
\begin{align}
    \frac{1}{m}\left[p_\tau+u_x\right] + \left[p_T + up_x\right] &= 0 \\
    \frac{1}{m}\left[u_\tau+p_x\right] + \left[u_T+uu_x-ru\right] &= 0
\end{align}

The $\mathcal{O}(\frac{1}{m})$ terms give the canonical non-dimensional wave equation for $p$ or $u$; by cross differentiation and subtraction
\begin{equation}
    p_{\tau \tau}=p_{xx} \qquad \mbox{or}
    \qquad u_{\tau \tau}=u_{xx}
\end{equation}
Converting back to dimensional variables gives the canonical dimensional wave equations
\begin{equation}
    P_{tt}=c^2 P_{xx} \qquad \mbox{or}
    \qquad U_{tt}=c^2 U_{xx}
\end{equation}

The $\mathcal{O}$
(1) terms can be written using the non-dimensional convective  derivative in the slow time $T$, $\frac{D}{DT} = \frac{\partial}{\partial T} + u\frac{\partial}{\partial x}$
\begin{equation}
    \frac{Dp}{DT} = 0 , \qquad \qquad
    \frac{Du}{DT} = -ru
\end{equation}
Converting back to dimensional variables these equations for the slow time dynamics are
\begin{equation}
    \frac{DP}{DT} = 0 , \qquad \qquad
    \frac{DU}{DT} = -\mathcal{R}U
\end{equation}
where the dimensional convective derivative is $\frac{D}{Dt} = \frac{\partial}{\partial t} +U\frac{\partial}{\partial x}$. The boundary conditions for these equations are determined by the solution of the wave equation on the fast time scale.

These equations relate to the dynamics in individual arteries, a fast time scale related to wave travel and a slow time scale where convection dominates. To relate these results to the arterial tree we have to consider the connectivity of the arteries.

\subsection{Relating results in uniform tubes to the arterial tree}

The solution of differential equations depends not only on the equations but also the boundary and initial conditions. We have ignored this problem in the previous section by looking only at the generic nature of the solutions in a uniform elastic tube without any reference to the boundary conditions. Applying appropriate boundary conditions to arteries is difficult because of the anatomical complexity of the arterial system. The arterial system can best be described as a binary tree structure with occasional loops (e.g. the Circle of Willis in the cerebral circulation, the palmar and plantar arches in the hand and foot and the highly looped arterial arcades supplying the intestines). The total number of arteries in the circulation depends on the definition of the boundary between the arteries and the capillary circulation, but it is certainly a very large number. 

The total number of capillaries in the human body can be estimated in two ways: by the average measured density of capillaries in various tissues or by average capillary flow rates compared to cardiac output. Both methods give roughly the same estimate, 10 billion capillaries ($10 \times 10^9$) in man. Making the anatomically conservative estimate that each terminal artery perfuses at least 10 capillaries, \cite{Delashaw1988} there are roughly $10^9$ terminal arteries. In a  binary tree with N terminal branches there are N-1 supporting branches which means that the total number of arteries is estimated to be $2\times 10^9 -1$. In an arterial tree with loops, the estimate is greater because the arteries in the loops do not support terminal branches.

Two billion arteries precludes any exact modelling of the arterial system with current computer technology. There are two viable alternatives; statistical methods and integrative methods. Statistical methods are attractive and worth exploring but they will rely on extensive measurements and modelling to produce clinically useful results. We will use an integrative approach in this paper.

The conservation of mass, particularly for incompressible fluids, is most amenable to the integrative approach because the rate of change of a control volume is related directly to the net flux into the volume. We can exploit this to study the dynamics of the whole arterial volume.

\section{The wave solution}

The solution of the wave equation on the arterial tree is difficult because it depends upon detailed, clinically unavailable knowledge about the properties of the entire tree. Taking advantage of the two-time analysis we need only to assume that there is a solution and that it occurs on a fast time scale associated with a characteristic wave speed $c^*$. As the waves propagate through the arterial tree at the fast time scale they leave behind them changes in pressure and flow that act as boundary conditions on the slow-time behaviour of the arterial tree. We assume that the resultant pressure is the reservoir pressure $P_r(T)$, where $T=t$ is the slow time scale.

\subsection{Wave travel on the arterial tree}

We model the arterial system as a binary tree with 1-D elastic arteries (edges) connected to bifurcations (internal nodes) or to terminal resistances representing the resistance of the local microcirculation (terminal nodes). We assume the nodes are discontinuities and ensure the continuity of mass and momentum by applying the Kirchoff conditions: 1) the net flow into any node is zero and 2) pressure is constant across the node.

The wave equation is linear and Fourier's theorem demonstrates that any periodic function can be decomposed into the summation of sinusoidal waves at the harmonic frequencies with appropriate magnitudes and phase. For this reason most studies of arterial hemodynamics use the Fourier sinusoidal waves as the basis for their analysis. This is the approach used by Westerhof and his colleagues to derive the input impedance in man.\cite{Murgo1980}

We will take an alternative approach, originally developed to study waves in gas dynamics,  that uses discrete wavefronts (wavelets) as the basis of the analysis.\cite{Courant1962} This approach has a number of advantages: any waveform including non-periodic waveforms can be synthesised using successive wavelets of the appropriate magnitude and it is conceptually easier to determine the reflection and transmission of the waves when they encounter the discontinuities at the nodes.

It is easy to see that any function $f(z)$ is a solution of the wave equation when $z=x \pm ct$ because by the chain rule for partial differentiation $f_{xx}=f''(z)$ and $f_{tt}=c^2f''(z)$ where $^\prime$ denotes the total derivative. For a wavelet, we chose the step function $f = \Delta P \;\mathfrak{U}$ where $\Delta P$ is a constant amplitude and $\mathfrak{U}$ is the unit step function. The change in pressure $\Delta P$ can be positive or negative. When $z=x-ct$ the wavelet is travelling in the forward direction and when $z=x+ct$, the backward direction. Making the Galilean transformation to travel with the wavelet, conservation of mass and momentum show that the magnitude of the change in pressure across the wave is related to the magnitude of the change in velocity
\begin{equation}
\label{waterhammer}
    \Delta P = \pm \rho c \Delta U
\end{equation}
where plus is for forward waves and minus, backward. These equations are generally referred to as the \textit{water hammer} (or Joukowsky) equations.

\subsection{Wave reflections in terminal arteries}

In our model, terminal arteries, denoted by $k \in K$, are terminated by terminal resistances $R_k$. These terminal resistances represent the lumped effect of the downstream resistances of the microcirculation. When a wavelet encounters a terminal resistance, conservation of mass and momentum requires that a wavelet is reflected with the reflection coefficient
\begin{equation}
    \Gamma_k = \frac{R_k-Z_k}{R_k+Z_k}
\end{equation}
where $Z=\frac{\rho c}{A}$ is the characteristic impedance of the vessel. When $R \gg Z$, $\Gamma \approx 1$ and when $R \ll Z$, $\Gamma \approx -1$. Since wavelets reaching the terminal nodes are travelling forward the reflected waves travel backwards.

\subsection{wave reflections at internal nodes}

In our model all internal nodes $i \in I$ are bifurcations where 3 vessels converge. If we denote the vessel containing the incident wavelet as $a$ and the other two daughter vessels as $b$ and $c$, similar conservation considerations require that there is a reflected wavelet with reflection coefficient
\begin{equation}
    \Gamma_a = \frac{Y_a - (Y_b+Y_c)}{Y_a+Y_b+Y_c}
\end{equation}
where $Y=1/Z$ is the characteristic admittance of the vessel. There are also transmitted waves in vessels $b$ and $c$ with transmission coefficients equal to $\Gamma_a + 1$. If the admittance of the incident vessel is equal to the net admittance of the other two vessels, $\Gamma_a = 0$,  the bifurcation is well-matched and the incident wave is not reflected and passes into the other two vessels unchanged. 

Addition of the reflection coefficients for all three vessels meeting at a bifurcation shows that $\Gamma_a+\Gamma_b+\Gamma_c = -1$. This means that if a bifurcation is well-matched in one direction it cannot be well-matched in either of the other directions. There are no systematic studies of impedance matching in small artery bifurcations (diameter <300 $\mu$m),  but it is usually observed that bifurcations in the large conduit arteries are nearly well-matched in the forward direction (defined as the net direction of blood flow from the heart to the periphery), this means that backward waves travelling from the periphery to the root cannot be well-matched. This asymmetry of wave travel in the forward and backward directions in the arterial tree has important implications for arterial hemodynamics.

\subsection{Wave travel in the arterial tree}

In principle, once we know the connectivity of the arterial tree, we now have everything needed to track the wavelets as they travel through the tree. In practice this approach to arterial hemodynamics is impractical because the number of wavelets increases exponentially with the number of nodes encountered by the wavelets; each wavelet produces three new wavelets at each internal bifurcation, one reflected and two transmitted wavelets. In a study of wavelet transmission in a 55 artery model of the main arteries, it was found that after each wavelet had encountered 20 nodes there were approximately $10^{80}$ wavelets in the model.\cite{Parker2017} (For comparison, this is very close to the current estimate for the number of atoms in the universe!\cite{universetoday}) It is necessary to find an alternative more approximate method which is computable. 

The two-time analysis of the 1-D conservation equations provides us with just such an approach. The wavelets propagate through the individual arteries on the fast time scale $\tau$ establishing the boundary conditions for the convective changes on the slow time scale $T$. While the local changes due to the wavelets on the fast time scale can be quite complex, we know that the conservation laws are also valid for the arterial system as a whole on the slow time scale.

\section{The reservoir pressure}

The two-time analysis of the conservation equations for an individual arterial segment on the slow-time scale $T$ are to $\mathcal{O}(m)$ 
\begin{equation}
    \frac{DP}{DT} = 0 , \qquad \qquad
    \frac{DU}{DT} = -\mathcal{R}(A)U
\end{equation}
where  $DP/DT$ refers to the convective derivative $D/DT = \partial/\partial T + U\partial/\partial x$ and $\mathcal{R}(A)$ is a resistance term that depends on the cross-sectional area $A$ of the vessel. A solution to these equations over the whole arterial tree is also a difficult problem. The Kirchoff conditions at each node, internal and external, can be applied to give the inlet and outlet boundary conditions for the equations on each vessel. 
Since the flow into the root of the aorta changes with time, the distribution of $P$ and $U$ will also be a function of time. They are, to $\mathcal{O}(m)$, the quasi-steady solutions corresponding to the instantaneous inlet and outlet boundary conditions.

The solution for the pressure distribution on a large tree is difficult, even in steady state. It requires complete knowledge of the connectivity of the tree and the diameter and length of each vessel (edge) in the network. Almost none of this information is available clinically. It is therefore necessary to make assumptions and approximations.

We assume that the arterial network can be described by a bifurcating tree; in network theory, we take this to be the minimal spanning tree. From estimates above, this tree contains approximately $2\times 10^9$ arteries (edges). Which is in keeping with some limited empircal data.\cite{Zamir1988} The \textit{Terminologia anatomica} includes approximately 1000 named arteries and almost everything we know about the connectivity and properties of arteries is limited to these larger arteries. Stated in another way, we are almost completely ignorant of the properties of 99.99999\% of the arteries that make up the arterial system.

Faced with this complexity and ignorance, we can still make reasonable estimates of pressure and flow in the arterial tree through the use of compartmental analysis; i.e. treating some fraction of the arterial system as a single compartment with a single pressure.
To do this we use the long-time conservation equations to divide the arteries into the \textit{inviscid} arteries where $\mathcal{R}$ can be neglected and the \textit{resistance} arteries where the resistance term $\mathcal{R}$ cannot be neglected.

The resistance of the smaller arteries can be estimated with reasonable accuracy using Poiseuille's law; the Reynolds and Womersley numbers being sufficiently small. For Poiseuille flow the resistance coefficient is \begin{equation*}
    \mathcal{R}_n = \frac{8\pi \mu}{\rho A_n}
\end{equation*}
If, for example, we take the diameter of the terminal arteries (arterioles) to be $D_k=10\;\mu m$ and the diameter of the smallest of the inviscid arteries to be $D_i=100\; \mu m$, then $\frac{\mathcal{R}_i}{\mathcal{R}_k} = \frac{A_k}{A_i} = 0.01$, which can be considered to be negligible for our purposes.

For the inviscid arteries the long-time conservation equations are satisfied by a uniform pressure in each vessel which, because of the continuity of pressure Kirchov conditions at each bifurcation, extends throughout the inviscid arteries. It is important to note that this uniform pressure can change in time because of the variation of the changes in the flow into the root of the arterial tree. Physically these changes in the inlet boundary conditions give rise to waves that propagate on the fast-time scale establishing a new uniform pressure in the inviscid arteries at the slow-time scale.

The slow-time conservation equations also give a uniform mean velocity in each inviscid vessel. The velocity, however, is not uniform throughout the inviscid part of the tree because the Kirchov conditions at the bifurcations requires that the net flow into the node is zero; i.e. the flow in the parent vessel is equal to the sum of the flows in the two daughter vessels. The split of flow between the two daughters is determined by the net resistance downstream of each daughter including all of the downstream resistance arteries and the capillary circulation. The calculation of these resistances requires information that is not available clinically. 

The local flow in the arteries is also complicated by the elastic nature of the arterial walls. Mass conservation requires that the rate of change of the volume of the vessel is equal to the difference between the inflow and outflow volume flow rates. There is no steady state in the highly pulsatile arterial flow but if the flow is periodic mass conservation requires that the net flow out of each vessel over one period must equal the net flow into it. 

We are now in a position to derive useful information about the dynamic behaviour of the arterial system by defining a control volume $V$ defined by the lumen volumes of the inviscid arteries. From the slow time conservation equations we know that the pressure within $V$ is uniform but dependent on time. We define this pressure as the \textit{reservoir} pressure $P_r(t)$. 

The inlet to this control volume is the inlet to the aortic root and we assume that the volume flow rate $Q_0(t)$ is known. The outlets of the control volume are the outlets of the terminal inviscid arteries. Formally these can be defined by choosing some arbitrary terminal diameter (such as $100\; \mu m$ discussed above) and selecting the terminal artery as the smallest artery on every path through the arterial tree with $D_j\ge D_k$, i.e. $(min(D_j;D\ge D_k,j)$. To avoid confusion between the terminal inviscid arteries and the terminal resistance arteries, 
we will call the latter 'arterioles' and the former simply 'terminal arteries'. Conveniently, there is a unique path from the root to each edge in a tree which ensures that these terminal edges span the entire tree.

The conservation of mass can be expressed by the ordinary differential equation 
\begin{equation}
\label{dVdt}
    \frac{dV}{dt} = Q_0(t) - Q_{out}(t)
\end{equation}
where  $Q_{out}(t)$ is the time-dependent net volume flow rate out of the large number of terminal arteries.

To continue the analysis it is necessary to make two simplifying assumptions. The first is that the relationship between volume $V$ and reservoir  pressure $P_r$ is given by a linear compliance relationship
\begin{equation}
     \frac{dV}{dP_r} = C
\end{equation}
We will take $C$ to be a constant over the normal range of diastolic to systolic pressure. This is equivalent to the concept of incremental elasticity in solid mechanics where the elastic constant is considered to be constant over some increment of strain, not the entire range of possible strains. Integrating from the diastolic pressure $P_d$ we obtain the linear relationship 
\begin{equation}
  V(t) - V_d = C\left(P_r(t)-P_d\right)
\end{equation}
where $V_d$ is the volume of the arteries at the diastolic pressure $P_d$.

The second assumption is that $Q_{out}$ is related to $P_r$ through an Ohmic resistance $R$ over the range of pressure during the cardiac cycle 
\begin{equation}
    Q_{out} - Q_d = \frac{P_r-P_d}{R}
    = \frac{V-V_d}{RC}
\end{equation}
where $Q_d$ is the flow out of the arteries at the diastolic pressure $P_d$ and we have used the compliance relationship to replace pressure with volume. Physiologically, $P_d$ is greater than the venous pressure and so we will only consider cases where $Q_d > 0$, i.e. $Q_{out} > 0$ over the whole cardiac cycle.

Substituting, we can write the volume conservation equation as an ODE for the arterial volume $V(t)$
\begin{equation}
  \frac{dV}{dt} + \frac{V - V_d}{RC} = Q - Q_d
\end{equation}
We recognise $RC$ as the diastolic time constant (usually called $\tau$) from classical Windkessel theory but it is convenient to retain the product $RC$ in our analysis (and to avoid confusion with the notation for the fast time scale). For constant $RC$ this equation can be solved by quadrature for any $Q(t)$
\begin{equation} 
    V(t) - V_d = e^{-t/RC} \int\limits_0^t e^{t'/RC} Q(t') dt' - RCQ_d \left(1 - e^{-t/RC}\right)
\end{equation}

For the following analysis it is convenient to define the integral in this equation as the 'buffered' volume change due to the inflow, $V_Q$
\begin{equation} 
    V_Q(t) \equiv e^{-t/RC} \int\limits_0^t e^{t'/RC} Q(t') dt'
\end{equation}
If $Q$ is positive (i.e. we do not allow for retrograde flow back through the aortic valve), then $V_Q$ is a monotonically increasing function that can be calculated from $Q(t)$ if the time constant $RC$ is known. In terms of $V_Q$ the instantaneous volume is
\begin{equation} 
    V(t) - V_d = V_Q(t) - RCQ_d \left(1 - e^{-t/RC}\right)
\end{equation}

The physical interpretation of this equation is that the change in volume of the arteries from the volume at diastolic pressure is equal to the buffered volume change due to the flow rate from the ventricle minus an exponential term that represents the relaxation of the system towards the volume that would eventually be achieved if the heart ceased pumping. If $R$ and $C$ were absolutely constant then this expression could be written more conveniently in terms of the volume at zero pressure. This is a much stronger limitation on the theory than our 'incremental' assumptions that $R$ and $C$ are constant only over the pressure range from diastolic to systolic pressure that is experienced by the arteries during normal physiological conditions. 

\subsection{Periodicity}

So far the analysis is valid for any arbitrary input flow rate. Under normal physiological conditions the cardiac cycle is quasi-periodic and so we assume that $Q$ is periodic with cardiac period $T_c$. Taking $t=0$ to be the time of the end of diastole/start of systole, Periodicity requires that $V(T_c) = V(0) \equiv V_d$ which requires
\begin{equation}
  Q_d = \frac{V_Q(T_c)}{RC\left(1 - e^{-T_c/RC}\right)}
\end{equation}
Substituting this into the general solution gives an equation for $V-V_d$ as a function of $Q$, $RC$ and the cardiac period $T_c$
\begin{equation} 
    V(t) - V_d = V_Q(t) - V_Q(T_c)\left(\frac{1 - e^{-t/RC}}{1 - e^{-T_c/RC}}\right)
\end{equation}

\subsection{Diastole}

By the nature of the cardiac cycle, the cardiac period in the arteries can be divided into systole when the aortic valve is open and $Q>0$ and diastole, when the aortic valve is closed and $Q=0$. Defining the time of the end of systole as $T_s$, zero inflow during diastole means that $V_Q(t) = V_Q(T_s)$ during diastole. Thus, during diastole $T_s \le t \le T_c$
\begin{equation}
  V(t)-V_d = V_Q(T_s)\left(\frac{e^{-t/RC} - e^{-T_c/RC}}{1 - e^{-T_c/RC}}\right) 
\end{equation}
Evaluating this at the start of diastole ($t = T_s$) and using this to eliminate $V_Q(T_s)$, we get a simple exponential relationship for the volume during diastole in terms of the volume at the start of diastole $V_s=V(T_s)$
\begin{equation}
  V(t)-V_d = (V_s-V_d)\left(\frac{e^{-t/RC} - e^{-T_c/RC}}{e^{-T_s/RC} - e^{-T_c/RC}}\right) 
\end{equation}
From the linearity of the compliance equation, the volume can be written in terms of the pressure
\begin{equation}
  P_r(t)-P_d = (P_{es}-P_d)\left(\frac{e^{-t/RC} - e^{-T_c/RC}}{e^{-T_s/RC} - e^{-T_c/RC}}\right) 
\end{equation}
where we have written the pressure at the end of systole as $P_r(T_s)=P_{es}$ because $P_s$ is generally used to indicate the systolic pressure, i.e. the maximum pressure during systole.

This exponential decay of the pressure is in a slightly different format from the usual expression for the diastolic decay of pressure which is generally written in terms of the asymptote as $t \rightarrow \infty$. Since we are referencing everything to the conditions at the end of diastole, this expression is normalised so that $P(T_d)=P_d$ and there is no need to define the asymptotic pressure $P_\infty$, which can be difficult to determine experimentally and for which the assumption of constant $RC$  is dubious.\cite{Behnam2019}

Looking at the properties of the homogeneous pressure $P_r$ obtained from the ODE describing overall mass conservation in the inviscid arteries, we see that it satisfies all of the conditions of the reservoir pressure defined in the reservoir-wave hypothesis: 1) it is uniform throughout the inviscid arteries, 2) during diastole it predicts the exponentially falling pressure that is observed in the measured pressure and 3) it is periodic. During systole, when there is flow into the aortic root, it is not equal to the measured pressure because there is a time delay in $P_r$ due to the elastic nature of the inviscid arteries. This is discussed further in the following section.

\subsection{The physical interpretation of the reservoir pressure}

Like mass and momentum, energy must be conserved in a control volume. Energy conservation, however, is not straightforward because energy can take many forms. For example, viscous effects in fluids, like friction in solid body interactions, can convert mechanical energy into heat. This form of energy conversion is very difficult to measure directly and in most cases is deduced by assuming that energy is conserved. Energy conservation is, however, useful in understanding the physical nature of the reservoir pressure.

The pressure of a fluid is a measure of its potential energy; pressure has the units of energy per unit volume. The hydraulic work done by a fluid is equal to pressure times the volume flow rate so that the instantaneous hydraulic work done on the arterial system at the aortic root is $P_0Q_0$. In the inviscid arteries the viscous effects are negligible and so the work done on the elastic walls of the arteries must equal the difference between the hydraulic work at the aortic inlet and the net hydraulic work at the outlets of the terminal vessels. The reservoir pressure which we identify with the elastic energy in the walls of the inviscid arteries is the integral of the elastic work. When the work at the aortic inlet exceeds the work at the outlets, $P_r$ increases; when the work at the inlet is less than the work at the outlets, $P_r$ decreases.

During diastole when $Q=0$, mass conservation requires that the net volume outflow rate is equal to the rate of change of the volume of the inviscid arteries. Similarly, energy conservation requires that the net hydraulic work at the outlets of the terminal arteries is equal to the rate of change of the elastic energy of the walls of the inviscid arteries which is equal to the rate of change of $P_r$. If the compliance of the inviscid arteries is constant this leads to an exponential fall in $P_r$ with the rate constant $\frac{1}{RC}$.

The reservoir pressure $P_r$ is spatially uniform throughout the inviscid arteries unlike the measured pressure $P$ which is different at different sites in the arterial system. The excess pressure $P_e=P-P_r$ can be useful in explaining local dynamics at different sites.

When $P$ is measured in the aortic root, $P_e$ gives us information about the interaction between the left ventricle and the aorta. At the end of diastole $Q_0=0$ and $P=P_r$ so that $P_e=0$. During the iso-volumic contraction phase the left ventricular pressure increases rapidly. When it exceeds the pressure in the aortic root the aortic valve opens and the boundary conditions at the aortic root suddenly change to the continuity of pressure and flow between the left ventricle and the aortic root. These boundary conditions are satisfied dynamically by the production of forward compression wavelets that propagate very rapidly down the aorta. These waves dominate the local flow but there are still backward wavelets arriving at the aortic root which are generated by the falling pressure at the outlets of the terminal arteries.

The mass conservation equation for the inviscid arteries (\ref{dVdt}) says that the extrema of $V$, and hence $P_r$, will occur when $Q_0=Q_{out}$. This happens twice during the cardiac cycle. The first time is very early in arterial systole when $Q_0$ increases to equal  $Q_{out}$ which is falling but still substantial throughout diastole. This time corresponds to the minimum in $P_r$. This brief period of falling $P_r$ at the start of systole is a characteristic feature of the $P_r$ waveform and provides evidence that $P_r$ is not simply related to the backward wave calculated using Fourier analysis. The second time occurs late in arterial systole when the rate of ejection from the left ventricle falls to equal $Q_{out}$. This time corresponds to the maximum in $P_r$ which occurs before the closure of the aortic valve.

$P_e$ at the aortic root could be very important clinically if the forward compression wavelets produced by the contraction of the left ventricle during early arterial systole are so large that the backward wavelets at that time are negligible. If this is true, the water hammer equation (\ref{waterhammer}) for forward waves implies that $P_e$ is proportional to $Q_0$. This suggestion has received experimental support from detailed measurements in open-chest dogs \cite{Tyberg2005} and invasive and non-invasive measurements in humans.  \cite{Hughes2020} \cite{Armstrong2021} This assumption also provides the theoretical basis for the 'pressure-only' method for calculating $P_r$ from the measured $P$ waveform. \cite{Parker2009}.\cite{Hughes2020}.

When pressure is measured at some other site, $P_r$ and $P_e$ could also be useful in determining the local dynamics. For example, the pressure waveform can be  measured non-invasively from the radial artery and it has been observed that the waveforms differ between young and old individuals  complicating comparisons.\cite{O'Rourke2007} Understanding the dynamics of the radial pulse is complicated firstly because the radial artery is very close to the terminal arteries and secondly because the anastomosis of the radial and ulnar arteries through the palmar arteries could give rise to forward wavelets in the ulnar artery appearing as backward wavelets in the radial artery. Providing answers to questions about local dynamics in peripheral arteries will be challenging but it is likely that $P_r$ and $P_e$ will be involved.

\section{Discussion and Conclusions}

Two-time asymptotic analysis provides a framework for analysing arterial hemodynamics. It assumes that time $t$ can be separated into a fast-time $\tau$, dependent upon a characteristic wave speed $c^*$, and a slow-time $T$, determined by the characteristic velocity of the blood $U^*$. The ratio of these characteristic times is the Mach number $m = U^*/c^*$, which is generally small for arterial flows. In some young individuals $m$ may not be as small, since arterial wave speeds are lower, which suggests that the reservoir hypothesis should be used with due caution in the young.

Applying the two-time analysis to the 1-D conservation equations for mass and momentum shows that to $\mathcal{O}(m)$ the solution for the pressure and velocity obeys the wave equation on the fast-time scale and simple relationships involving the convective derivative on the slow-time scale. The wave equation is hyperbolic in nature and solutions depend only on the boundary conditions at the inlet and outlets of the arterial system and matching conditions at discontinuities such as bifurcations within the arterial network. The fast-time solutions dictate the boundary conditions at the slow-time scale. 

The slow-time conservation equations involve a resistance term (the viscous resistance coefficient $\mathcal{R}$) and lead to the suggestion that the arteries can be divided into inviscid arteries where viscous effects can be neglected and resistance arteries where viscous losses are important. This assumption is similar in principle to the  boundary layer approximation in aerodynamics. In the inviscid arteries the long-time conservation equations and the Kirchoff conditions at the bifurcations show that the pressure is homogeneous in space and variable in time. We identify this pressure with the reservoir pressure $P_r$ originally introduced in the heuristic reservoir-wave hypothesis.

We apply overall mass conservation to the inviscid arteries to calculate the dynamics of $P_r$ as a function of the known volume flow rate into the aortic root from the left ventricle $Q_0(t)$, the net resistance of the resistance arteries and the microcirculation $R$ and the net compliance of the inviscid arteries $C$.

The physical meaning of the reservoir pressure, revealed by the theoretical derivation here, is straightforward: it is the potential energy stored by the elastic walls of the inviscid arteries during the cardiac cycle. The energy stored during the early part of systole is recovered during late systole and, particularly, during diastole when it generates flow through the resistance arteries and microcirculation, smoothing out the highly pulsatile flow generated by the ventricle. This corresponds to the commonly held view of the 'arterial reservoir' by many physiologists and cardiologists and explains why the Windkessel has remained as a useful teaching concept. This interpretation is also supported by an earlier analysis using the calculus of variation to show that the hydraulic work done by the reservoir pressure is the minimum work necessary to generate any given flow rate in the arteries.\cite{Parker2015}.

There have been a number of papers that have been  critical of the reservoir-wave hypothesis.\cite{Westerhof2015}\cite{Mynard2014} We believe these criticisms are a result of a misunderstanding about the role of waves in generating the reservoir pressure. The two-time analysis that is the basis of our theoretical approach should clarify these misunderstandings and lead to a wider acceptance of the reservoir pressure as a useful concept in arterial hemodynamics. 

The first misunderstanding by the critics of reservoir pressure is the belief that $P_r$ is directly related to the backward wave $P_b$ that is obtained when the measured $P$ is separated into its forward and backward components.\cite{Westerhof2015} It is true that $P_r \approx 2 P_b$ during diastole when $Q=0$ and hence $P_f=P_b$ but it is not true during systole when $Q \ne 0$, and $P_r$ calculated using the methods discussed above is generally different from $2P_b$.

The second misunderstanding about reservoir pressure comes from 1-D computational studies of arterial models where the first waves returning to the aortic root deviate from the reservoir pressure.\cite{Mynard2014} The two-time analysis shows that the reservoir pressure results from the boundary conditions established by all of the waves in the arterial system, not just the early reflected waves. In fact, it is likely that some of the wavelets determining the reservoir pressure are very old, having been generated by ventricular contractions several beats before the current beat.\cite{Willemet2015} These wavelets are very small in magnitude due to the large number of reflections and re-reflections they have experienced, but there is an enormous number of them, and it is not obvious that they can be neglected. It is clear, however, that $P_r$ is not the resultant of the first few reflections that are calculated by computational modelling.

This study gives the reservoir pressure a rigorous theoretical basis for the first time. The two-time analysis is an approximate asymptotic analysis but it is mathematically rigorous and gives us a quantitative bound on the approximations that have been made; the results are accurate to $O(m^2)$, where $m$ is the characteristic Mach number for arterial flows. Since arterial wave speeds are typically at least an order of magnitude greater than the mean velocity, the results should be accurate to within a few percent. This is commensurate with the accuracy of other studies of biological flows, experimental and theoretical.

Being theoretically rigorous, of course, does not ensure that the results of the analysis are useful. That question can only be answered by clinical studies and practice. The theoretical results do however give us confidence in the validity of the concept of separating measured arterial pressure into reservoir pressure and excess pressure. It also gives us insight into the physical basis of the reservoir pressure which may be useful in understanding future clinical results.

\end{document}